# Beyond Authorship: Analyzing Contributions in *PLOS ONE* and the Challenges of Appropriate Attribution

Abdelghani Maddi[1*], Jaime A. Teixeira da Silva[2]

[1] *abdelghani.maddi@cnrs.fr*
ORCID : https ://orcid.org/0000-0001-9268-8022
GEMASS – CNRS – Sorbonne Université, 59/61 rue Pouchet 75017 Paris, France.

[2] *jaimetex@yahoo.com*
ORCID : https ://orcid.org/0000-0003-3299-2772
Independent researcher, Ikenobe 3011-2, Kagawa-ken, 761-0799, Japan.

* Corresponding author

**Abstract**

*Purpose:*
This study aims to evaluate the accuracy of authorship attributions in scientific publications, focusing on the fairness and precision of individual contributions within academic works.

*Design/Methodology/Approach:*
The study analyzes 81,823 publications from the journal *PLOS ONE*, covering the period from January 2018 to June 2023. It examines the authorship attributions within these publications to try and determine the prevalence of inappropriate authorship. It also investigates the demographic and professional profiles of affected authors, exploring trends and potential factors contributing to inaccuracies in authorship.

*Findings:*
Surprisingly, 9.14% of articles feature at least one author with inappropriate authorship, affecting over 14,000 individuals (2.56% of the sample). Inappropriate authorship is more concentrated in Asia, Africa, and specific European countries like Italy. Established researchers with significant publication records and those affiliated with companies or nonprofits show higher instances of potential monetary authorship.

*Research Limitations:*
Our findings are based on contributions as declared by the authors, which implies a degree of trust in their transparency. However, this reliance on self-reporting may introduce biases or inaccuracies into the dataset. Further research could employ additional verification methods to enhance the reliability of the findings.

*Practical Implications:*
These findings have significant implications for journal publishers, highlighting the necessity for robust control mechanisms to ensure the integrity of authorship attributions. Moreover, researchers must exercise discernment in determining when to acknowledge a contributor and when to include them in the author list. Addressing these issues is crucial for maintaining the credibility and fairness of academic publications.

*Originality/Value:*
This study contributes to an understanding of critical issues within academic authorship, shedding light on the prevalence and impact of inappropriate authorship attributions. By calling for a nuanced approach to ensure accurate credit is given where it is due, the study underscores the importance of upholding ethical standards in scholarly publishing.

**Competing interests**

The authors have no relevant financial or non-financial interests to disclose.

**Authors contributions**
Abdelghani Maddi: conceptualization; data curation; formal analysis; investigation; methodology; project administration; supervision; validation; visualization; writing – original draft preparation; writing – review and editing.
Jaime A. Teixeira da Silva: conceptualization; investigation; methodology; supervision; validation; visualization; writing – review and editing.
**Data availability**
The code to generate the data is available on Github at: https://github.com/abdelghani-maddi/PLOS_ONE_authorship

The data are available in Zenodo: https://zenodo.org/records/10568892

**Funding information**

The authors received no specific funding for this work.

# 1. Introduction

Scientific publications fulfill a critical role in disseminating research findings and advancing knowledge across various disciplines. Within this scholarly communication process, authorship holds significant importance as it represents not only credit and recognition, but also responsibility and accountability (Kerr et al., 2018; Smith & Williams-Jones, 2012). However, the traditional concept of authorship often falls short in accurately capturing the diverse contributions made by individuals in a research project, thereby creating challenges in assessing and acknowledging their roles appropriately. Additionally, concerns regarding scientific integrity and transparency have arisen due to issues such as guest, gift, or honorific authorship (Brand, 2012).

To address these challenges, frameworks like the Contributor Roles Taxonomy (CRediT) were developed (Allen et al., 2019; Holcombe et al., 2020). CRediT offers a standardized approach to describing and acknowledging researchers' specific contributions in scholarly publications, providing a comprehensive set of predefined roles that extend beyond the conventional notion of authorship, encompassing various aspects of the research process. These roles include conceptualization, methodology, data curation, writing (original draft and review), and more.

The implementation of clear authorship criteria and adoption of frameworks like CRediT are crucial for promoting scientific integrity, enhancing transparency, accountability, and fairness in attributing authorship, as well as mitigating risks associated with inappropriate authorship practices (Kerr et al., 2018; McNutt et al., 2018). While these frameworks offer structured approaches, their real-world verification poses challenges, as editors often struggle to effectively verify contributions' accuracy, and issue that raises concerns about the practical implementation of such criteria (Teixeira da Silva, 2023). Until May 2023, Public Library of Science (PLOS) journals, adhered to strict authorship criteria recommended by the International Committee of Medical Journal Editors (ICMJE). However, since May 2023, PLOS journals have adopted the recommendations of McNutt et al. (2018), advocating for the use of the CRediT taxonomy, which provides precise definitions of authorship. While both frameworks aim to clarify authors' contributions, the ICMJE focuses on general criteria, while CRediT offers a more detailed taxonomy of specific contributions that each author may have made. However, both frameworks share the goal of ensuring that only those who have made a substantial contribution to the research are listed as authors. These criteria emphasize significant contributions, stipulating that mere funding or resource provision does not warrant authorship. Instead, individuals involved in project administration or resource provision should be acknowledged separately.

The issue of merit and recognition of author contributions within research teams is a contentious one, and can lead to conflicts within the academic community (Savchenko & Rosenfeld, 2024; Smith et al., 2020). In their survey of more than 8,000 researchers, Smith et al. (2020) showed that nearly half of respondents had encountered reported disagreements regarding authorship naming at least once in their career. Furthermore, a recent international and interdisciplinary survey conducted by Savchenko and Rosenfeld (2024) with 752 scholars from 41 research fields and 93 countries revealed that conflicts regarding authorship credit allocation often arise early in academic careers. Specifically, nearly one in four participants reported experiencing at least one conflict with an advisor during their master's or doctoral studies. Furthermore, the issue of nepotism in academic authorship is a subject of concern, so authors should declare spousal and kinship relationships within the same author team to prevent potential conflicts of interest (Teixeira da Silva & Rivera, 2021).

In this article, we aimed to examine whether there are articles published in *PLOS ONE* that do not adhere to ICMJE and CRediT criteria contribution criteria, and consequently, authors who should not have been included in the list of authors. It is worth noting that we selected *PLOS ONE* for our analysis, despite other options like *PLOS Medicine*, due to its significantly larger volume of publications, which provided a substantial sample for our investigation. To achieve this, we analyzed 91,626 publications from *PLOS ONE* published between January 2018 and June 2023. We extracted the authors' contributions and their affiliations.

## 2. Overview of recent literature

There is a substantial body of literature on authorship and author contributions, and this brief literature review focuses on the issues surrounding the definition of author roles in publications.

The question of defining roles within research teams is crucial, particularly in the context of the significant increase in "big team science" projects over the past decade (Baumgartner et al., 2023). In such projects, a poorly defined or ambiguous allocation of roles for each contributor can be problematic for project management, beyond the issue of symbolic credit allocation. Additionally, Brand et al. (2015) identified several benefits of having a clear definition of author roles. For example, it provides important information to policymakers and research funders regarding team composition. That is why the "co" status on a paper is so vital, because it can be rewarded (Teixeira da Silva, 2021). It can also be valuable for researchers themselves when seeking colleagues to complement ongoing projects, particularly technical aspects of them. Furthermore, it can aid journals in identifying potential article reviewers based on their expertise as defined by their roles in scientific publications (Brand et al., 2015).

Beyond the managerial and practical aspects of defining authors' roles, it is worth noting that one of the major challenges is addressing questionable practices in this domain. In the realm of scientific publishing, inappropriate authorship allocation reflects complex ethical dilemmas and deeply rooted systemic issues within academia. Beyond the examples provided, a deeper exploration of the underlying dynamics and their implications is essential to grasp the full extent of the problem.

For instance, authorship allocation based on financial incentives extends far beyond cases where individuals directly pay to be listed as authors (Abalkina, 2023; Abalkina & Bishop, 2023). In some contexts, researchers may be incentivized to include influential figures or funders as authors to secure future funding or expand their professional network. This practice, often termed "gift authorship," blurs the line between genuine intellectual contribution and strategic maneuvering (Lissoni & Montobbio, 2015; Teixeira da Silva & Dobránszki, 2016). Additionally, the commercialization of research through industrial partnerships or corporate funding introduces further complexities, with authors potentially prioritizing commercial interests over academic integrity. Collaborations with pharmaceutical laboratories or certain corporations serve as poignant examples of this phenomenon, where the pursuit of profit may overshadow scientific objectivity and ethical considerations (Sismondo, 2020).

Concurrently, authorship allocation based on access to research resources creates imbalances, especially in resource-constrained environments or institutions with varying levels of infrastructure (Kwiek, 2020). Individuals with privileged access to cutting-edge equipment or proprietary data may exert disproportionate influence on authorship decisions. Collaborators from affluent institutions may be included as authors based solely on their affiliation, regardless of their substantial contribution to the research (Morton et al., 2021). This perpetuates a cycle of inequality where researchers from underprivileged backgrounds struggle to compete on an equal footing.

Moreover, power dynamics within the academic community can exert undue pressure on authorship decisions, particularly when junior researchers feel compelled to include their supervisors or mentors as authors for fear of reprisals or jeopardizing their career prospects. Similarly, collaborators from prestigious institutions may leverage institutional prestige to negotiate authorship, irrespective of their actual involvement in the research. Such practices undermine the principles of academic meritocracy (Dusdal & Powell, 2021) and erode trust within research collaborations.

Furthermore, cultural and disciplinary norms shape perceptions of authorship allocation,

contributing to variations in ethical standards and practices. In some cultures, hierarchical structures or deference to authority may influence authorship decisions, leading to disproportionate credit allocation (Hoekman & Rake, 2024). Likewise, disciplinary norms regarding collaboration and acknowledgment vary, with some fields emphasizing individual contributions while others prioritize collective efforts (Whetstone & Moulaison-Sandy, 2020). These cultural and disciplinary variations underscore the complexity of addressing inappropriate authorship practices globally, necessitating nuanced approaches tailored to specific contexts.

Additionally, gender disparities prevalent in authorship practices (Bernardi et al., 2020; Ni et al., 2021) contribute to the complexity of the issue. Despite authorship being the primary form of symbolic capital in science, women often face challenges regarding fair attribution of credit. An international survey examining gendered practices in authorship communication, disagreement, and fairness found that women were more likely to experience authorship disagreements and have their contributions devalued (Ni et al., 2021). This disparity in perception extends to the acknowledgment of credit, with women feeling undervalued compared to men. Such systematic undervaluation of women's contributions in science perpetuates cumulative disadvantages in their scientific careers.

Consequently, inappropriate authorship allocation in scientific publications is a multifaceted phenomenon driven by financial, institutional, and sociocultural factors.

## 3. Data and Method

*3.1. Data extraction and processing*

The data extraction and enrichment procedure involved web scraping techniques, R-4.3.1 scripting, and integration with the OpenAlex database (https://openalex.org/). The data was extracted for the period from January 1, 2018 to June 6, 2023. The data are extracted for the past five and a half years, as during this period, efforts in disseminating information on best practices and scientific integrity have been pervasive within the scientific community, and all researchers should be aware of these issues. The script interacts with *PLOS ONE* web pages, extracts specific information, and integrates it with relevant metadata from OpenAlex for a comprehensive dataset.

The extraction and enrichment processes involved the following steps:
- Web page retrieval: The procedure retrieved web pages from *PLOS ONE* using appropriate search queries.

- HTML analysis: The web page's HTML structure was analyzed to identify the specific nodes or elements containing the required information, including author roles, DOIs, affiliations, and funding acknowledgments.
- Data extraction: Using web scraping techniques, the script extracted the desired information from the identified HTML nodes, ensuring consistency and accuracy.
- OpenAlex (Works entity) integration: The extracted data was enriched by integrating additional metadata from the OpenAlex database, including citations received by the articles and information about the authors.
- OpenAlex (Authors entity) integration: We also extracted the information relating to the authors from the "Authors" entity of OpenAlex. For example, authors' main affiliation addresses, affiliation types, number of works, number of citations, etc.
- Data transformation: The extracted and enriched data was cleaned, organized, and formatted (including with regard to the harmonization of author affiliations).

The scripts used for the automated extraction and enrichment process are openly available on the GitHub repository at the following URL: https://github.com/abdelghani-maddi/PLOS_ONE_authorship. Researchers and data analysts can access and utilize these scripts to replicate the extraction procedure, customize it for their specific needs, and contribute to its further development. For more details on the method, see Figure 1.

**Figure 1: Data extraction, cleaning, preparation and validation**

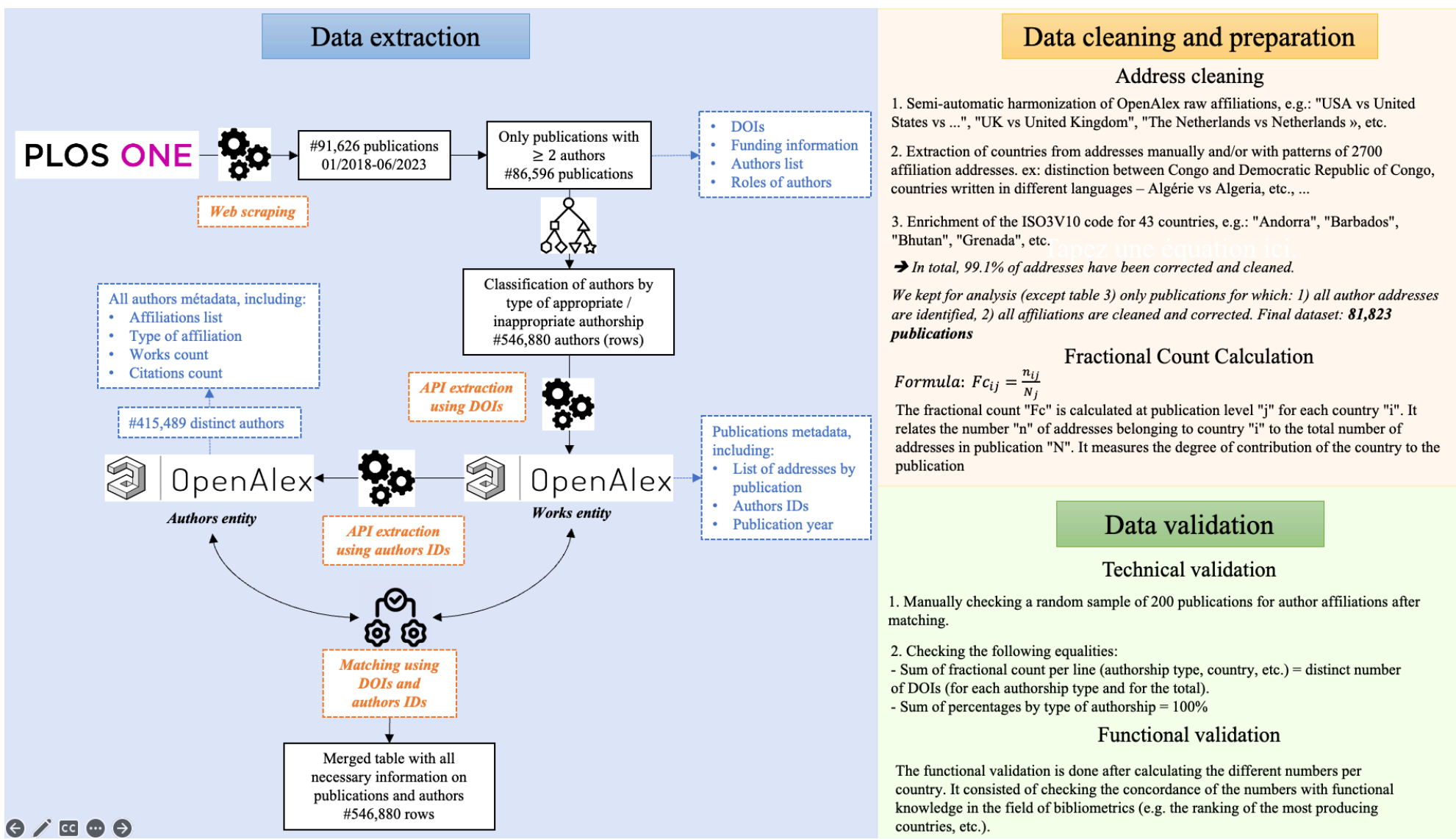

Please note that while the scripts provided on the GitHub repository were thoroughly tested, researchers should exercise caution and review the code before using it in their own projects, ensuring compatibility with their specific requirements and adhering to best practices in data extraction and analysis (Nichols et al., 2017).

3.3. Classification method of potential inappropriate authorship

In this study, we aimed to detect possible inappropriate authorship or violations of authorship criteria in *PLOS ONE* publications, with a focus on measurability and quantitative detection. Therefore, aspects related to the approval of the final version (condition 3 for authorship) and assuming accountability for all facets of the research (condition 4) are considered as met, even though this may lead to an underestimation of our calculations regarding inappropriate authorship. This is why we refer to eligibility for authorship in Table 1, provided that conditions 3 and 4 are met.

Table 1 outlines the 14 criteria of the CRediT taxonomy, accompanied by a column denoting whether each criterion, when met, is independently sufficient to confer authorship for an article. In refining our authorship definition, as highlighted previously, we hypothesize that authors have endorsed the final version (condition 3) and assume accountability for all facets of the research (condition 4). We adopt a more flexible stance on condition 2 (contribution to the article's drafting), recognizing that contributors need not adhere to overly stringent requirements in this aspect. According to our refined criteria, any contributor who has provided intellectual (substantive) input to the research, as delineated in Table 1, may be deemed deserving of authorship, contingent upon the fulfillment of conditions 3 and 4. This distribution is based on the definition provided by McNutt et al. (2018) and the discussion provided by Patience et al. (2019) on intellectual contributions deserving authorship.

**Table 1: CRediT criteria based on whether they are sufficient on their own for authorship or not**

| Contributor role | Role definition | Authorship eligibility |
|---|---|---|
| Conceptualization | Ideas; formulation or evolution of overarching research goals and aims. | Yes |
| Data Curation | Management activities to annotate (produce metadata), scrub data and maintain research data (including software code, where it is necessary for interpreting the data itself) for initial use and later reuse. | Yes |
| Formal Analysis | Application of statistical, mathematical, computational, or other formal techniques to analyze or synthesize study data. | Yes |
| Investigation | Conducting a research and investigation process, specifically performing the experiments, or data/evidence collection. | Yes |
| Methodology | Development or design of methodology; creation of models | Yes |
| Software | Programming, software development; designing computer programs; implementation of the computer code and supporting algorithms; testing of existing code components. | Yes |
| Validation | Verification, whether as a part of the activity or separate, of the overall replication/reproducibility of results/experiments and other research outputs. | Yes |
| Visualization | Preparation, creation and/or presentation of the published work, specifically visualization/data presentation. | Yes |
| Writing – Original Draft Preparation | Creation and/or presentation of the published work, specifically writing the initial draft (including substantive translation). | Yes |
| Writing – Review & Editing | Preparation, creation and/or presentation of the published work by those from the original research group, specifically critical review, commentary or revision – including pre- or post-publication stages. | Yes |
| Funding Acquisition | Acquisition of the financial support for the project leading to this publication. | No |
| Project Administration | Management and coordination responsibility for the research activity planning and execution. | No |
| Resources | Provision of study materials, reagents, materials, patients, laboratory samples, animals, instrumentation, computing resources, or other analysis tools. | No |
| Supervision | Oversight and leadership responsibility for the research activity planning and execution, including mentorship external to the core team. | No |

Therefore, based on Table 1, we identified three potential levels of potential inappropriate authorship, which are summarized in Figure 2. It is important to note that for a given publication, multiple types of inappropriate authorship may coexist. For instance, a publication involving collaboration among four authors may involve two authors who do not meet authorship criteria.

**Figure 2: Three levels of potential inappropriate authorship**

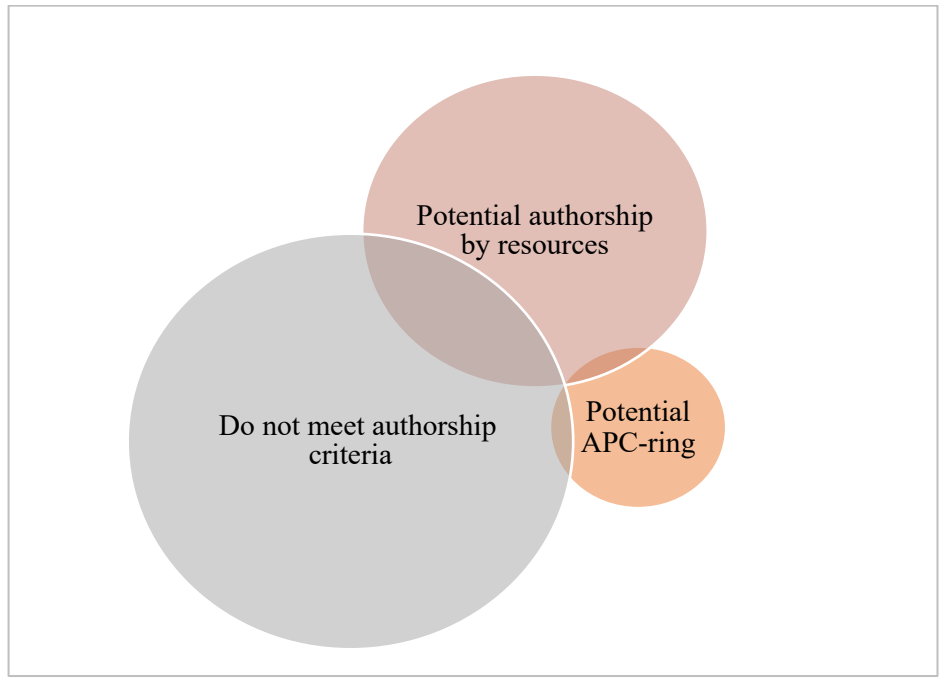

- Do not meet authorship criteria

These are publications in which at least one author does not meet the authorship criteria as defined by the journal *PLOS ONE*. In other words, these are publications in which at least one author does not have a substantial contribution to the paper in terms of design, analysis, or writing. These articles were identified by filtering authors who do not have any criteria among those tagged as "Yes" in Table 1.

- Potential authorship by resources

This is a subset of the previous type of publications, which refers to publications in which at least one "author" is considered as such solely because they have made a material and/or financial contribution (rather than an intellectual one). In the CRediT taxonomy, this corresponds to the roles of "Funding Acquisition" and "Resources." These publications may be funded by various funding sources.

- Potential APC-ring

This type is also a subset of the previous one and consists of publications that can be classified as Potential APC-ring publications as described by (Teixeira da Silva, 2024). These are publications that have no funding, but at least one author's sole role is to provide funding ("Funding Acquisition") that would be used to pay the *PLOS ONE* APC.

Based on this classification, we thus graded the three types of inappropriate authorship into three overlapping groups, of differing sizes (crude relative size indicated in Figure 1).

Furthermore, it is important to note that this study is based solely on the facts and statements provided by the authors and does not take into account possible misconduct related to authors who falsely claim roles they did not actually fulfill. Additionally, we are aware that the third type (Potential APC-ring) may be underestimated because there may be articles that fall under the "Authorship through by resources" category, where the author in question did not contribute to securing the funding but only paid the APC. It is worth noting that while CRediT provides a precise definition of resources, it does not explicitly include APCs. While there may not be explicit evidence or references in CRediT to support this, it is a plausible scenario that could exist in practice.

On the other hand, it is crucial to consider the possibility of overestimation in our figures, due to *PLOS ONE*'s guideline, which specifies that only corresponding authors are required to list authors' roles during submission. Importantly, we view this practice as an advantage rather than a limitation, as relying on corresponding authors increases the likelihood that genuine roles are accurately declared. *PLOS ONE* authors' guidelines ask the corresponding author to list "at least one criterion" from CRediT. The authors are aware that non-native English speakers might have misinterpreted this directive, potentially influencing the reported statistics.

### 3.4. Calculated indicators

#### 3.4.1. *Inappropriate authorship concentration index by country*

To represent the concentration of inappropriate authorship, we introduced an indicator termed the Potential Inappropriate Authorship Concentration Index by country/region (PIACI)." This index aims to correct for the size effect and facilitate cross-country comparisons. It is computed using the following formula:

$$IAI = \frac{S_{i_{inap}}}{S_{i_{PLOS}}}$$

The PIACI metric relates the country "i's" share within the corpus of publications with inappropriate authorship ($S_{i_{inap}}$) to its share within the total *PLOS ONE* dataset ($S_{i_{PLOS}}$). We utilized the fractional count to account for each country's contribution within each subset. For instance, if in a publication with inappropriate authorship, there are ten addresses from the United States and one from Italy, the United States will be assigned a weight of 10/11. This approach allows for a more precise calculation of the PIACI.

#### 3.4.2. *Specialization index in inappropriate authorship by affiliation type*

Specialization index (SI) in inappropriate authorship by affiliation type is constructed in a manner similar to the PIACI. It serves to gauge the "concentration" of each affiliation type within a specific category of inappropriate authorship. It is an index that also adjusts for size, with a neutral value of 1. It is computed using the following formula:

$$SI = \frac{S_{aff_{inap}}}{S_{aff_{PLOS}}}$$

The calculation involves relating the share of each affiliation type within a given inappropriate authorship category ($S_{aff_{inap}}$) to its share within the total *PLOS ONE* dataset ($S_{aff_{PLOS}}$).

## 5. Results

In this section, we present the key findings of our analysis. Firstly, we provide an overview of the global distribution of *PLOS ONE* publications by country from 2018 to 2023.

Figure 3 provides a comprehensive insight into the distribution of publications across different countries within the context of the *PLOS ONE* journal, employing a fractional counting methodology. Fractional counting is a valuable approach in bibliometric analysis as it takes into account the multiple affiliations that authors often have with institutions from various countries. This method, thus, allows for a more nuanced understanding of the global landscape of scientific contributions.

**Figure 3: Number of publications by country/region in *PLOS ONE* (01/2018-06/2023)**

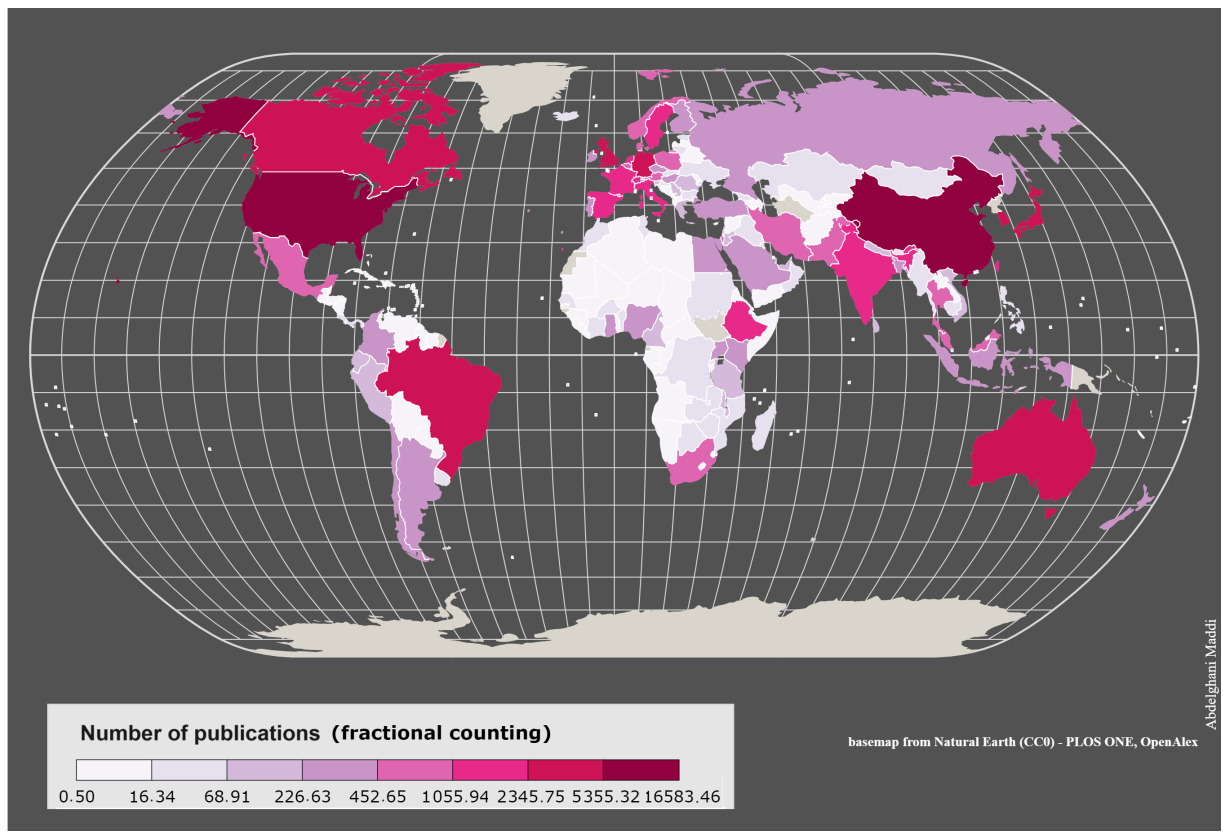

As expected, the figure reveals a dominance of research output from the world's major scientific hubs, with the United States and China leading the way. This is consistent with the well-established status of these countries as powerhouses in scientific research (Marginson & Xu, 2023). Following behind are nations such as Germany, the United Kingdom, Brazil, and Australia, all of which have strong traditions of scientific excellence.

What is particularly intriguing about the findings is the notable presence of Ethiopia in the publication distribution. Despite not being counted among the top 20 global knowledge producers (https://www.scimagojr.com/countryrank.php), Ethiopia emerges as a significant contributor to *PLOS ONE*. An unexpected surge in Ethiopia's publication count may be influenced, among various factors, by the exemption of APCs for the country, as outlined in the eligibility list (https://www.research4life.org/access/eligibility/). This observation raises intriguing questions, suggesting a potential surge in the scientific community within Ethiopia or collaborative efforts between Ethiopian researchers and their international counterparts.

Another aspect to consider is that, according to information available in the *PLOS ONE* FAQs[1] (https://plos.org/publish/publishing-faqs/), it appears that the APC exemption applies if all authors of a manuscript are members, with their publication fees automatically covered. Furthermore, if you are the corresponding author and your institution is a member, the FAQ currently indicates the waiving of the non-member fee.

This situation thus raises the possibility that Ethiopian researchers may benefit from the APC exemption even when collaborating with authors from high-income countries (HICs). This could potentially encourage practices such as recruiting academics from low-income countries (LICs) listed as eligible for a full APC waiver, ensuring that they are the corresponding authors.

As a result, while the APC exemption is a plausible explanation, a thorough investigation is needed to comprehensively understand the dynamics behind this phenomenon. This underscores the importance of ongoing monitoring and a deep understanding of publication policies to ensure the integrity of the scientific research process.

Overall, the distribution of publications presented in Figure 3 aligns with broader patterns observed in global scientific studies, showcasing the prominence of established research giants (see for example: https://www.scimagojr.com/countryrank.php) while also shedding light on unexpected yet noteworthy contributors, underscoring the dynamic and diverse nature of contemporary scientific research.

---

[1] See the question: "*If my institution is a member, will my publication fees be waived?*", and the corresponding answer: "*If all authors of a manuscript are members, their publication fees are automatically covered. If you are the corresponding author and your institution is a member, we are currently waiving the non-member fee.*"

Table 2 provides a breakdown of publications based on the type of potential inappropriate authorship observed within our analysis. It is important to note that for a given publication, multiple types of inappropriate authorship may coexist (The two examples below illustrate a scenario where a single author does not adhere to authorship criteria, and another example involving two authors with potential inappropriate authorship). Notably, the majority of publications, in 74,346 (90.86%) manuscripts, contributions listed justified authorship of all the authors listed. However, it is imperative to address the remaining subset of publications, which collectively constitute more than 9% of the dataset, totaling 7,477 publications.

Below are screenshots (taken from two randomly selected articles) illustrating two scenarios that do not adhere to the authorship criteria of *PLOS ONE*. In the first case, one of the authors falls under the category of "Potential APC Ring," while in the second case, two authors do not meet the authorship criteria, with each falling into different categories: "Potential Authorship by Resources" and "Does not meet PLOS ONE authorship criteria."

Example 1: Potential APC ring

Example 2: "Potential Authorship by Resources" and "Does not meet authorship criteria"

[Screenshot of a PLOS ONE article page showing "About the Authors" section with author roles and affiliations. Two authors are highlighted with red circles: one labeled "Potential authorship by resources" (ROLES: Resources) and another labeled "Not meet authorship criteria" (ROLES: Resources, Supervision).]

Table 2 provides a systematic breakdown of articles according to the contributions of the authors and the types of authorship they embody. It categorizes articles based on whether all authors adhere to *PLOS ONE* criteria or if one or more authors are classified under various categories of inappropriate authorship. For example, there is 585 articles with at least one author classified as "Not meet authorship criteria" and at least one author categorized as "Potential authorship by resources."

**Table 2: Number of DOI-based papers that fall into categories 1-3 of potential inappropriate authorship**

| Authorship type | Not meet authorship criteria | Potential authorship by resources | Potential APC ring | Count | # by type | % by type |
|---|---|---|---|---|---|---|
| Appropriate authorship | No | No | No | 74,346 | 74,346 | 90.86% |
| Inappropriate authorship | Yes | No | No | 4849 | 7477 | 9.14% |
| | No | Yes | No | 2008 | | |
| | Yes | Yes | No | 585 | | |
| | No | No | Yes | 21 | | |
| | Yes | No | Yes | 14 | | |
| Total | | | | 81,823 | 81,823 | 100.00% |

Of particular concern is a category akin to scientific misconduct, termed the "APC ring", pertaining to papers wherein at least one author's sole contribution is the payment of APCs. In this study, these instances may relate to papers where the authors did not receive institutional funding, and the APCs for *PLOS ONE* were covered by one of the authors who did not contribute to the research itself. It is essential to note that while these cases raise concerns, we do not have concrete evidence to conclusively categorize them as actual APC rings. Table 2 illuminates that while this particular scenario remains relatively rare, it is not altogether absent, affecting 35 articles in our dataset. Among these, 14 instances stand out, where there is at least one author responsible for financing the publication and at least one other author whose contribution does not align with the established criteria for authorship. These findings underscore the importance of ongoing education and awareness initiatives for authors who may be unfamiliar with such practices. Training programs and awareness campaigns could play a crucial role in ensuring that authors fully comprehend the significance of transparent and accurate authorship declarations, particularly in the context of evaluations and assessments within the academic community.

Furthermore, Table 2 sheds light on the category of "authorship by resources," which encompasses 2,593 publications, delineated into two distinct scenarios. Firstly, we identify 2,008 publications where at least one author's sole declared contribution is the provision of either funding, including institutional funding, or resources such as materials or equipment. Secondly, within this category, we observe an additional layer of complexity, involving publications where, in addition to the aforementioned scenario, there exists at least one other author whose sole role does not align with the established authorship criteria. For instance, their contribution may be centered around research supervision or administrative duties rather than direct research involvement. This dual-layered situation concerns 585 publications.

Finally, within Table 2, we observe a distinct category comprising 4,849 publications that deviate from *PLOS ONE*'s established authorship criteria, irrespective of financial and resource-related aspects. If the authors' declarations are true, these publications involve at least one author whose primary role revolves around administrative duties and/or research supervision. It is essential to emphasize that this category also encompasses authors who, in addition to assuming these roles, may contribute financially or provide resources to the research endeavor. While their involvement may be crucial for the smooth execution of the research project, including them in the author list would be considered inappropriate according to established authorship guidelines.

In these instances, it is pertinent to underline that these contributors should have been appropriately acknowledged in the acknowledgment section of the paper, recognizing their valuable support and assistance without conferring authorship status upon them (Teixeira da Silva et al., 2023). This distinction is vital in maintaining the integrity and transparency of authorship attribution, ensuring that those who make significant but non-research-oriented contributions are duly recognized while upholding the rigorous standards of authorship within the realm of academic publishing.

Table 3 provides a detailed distribution of authors categorized by different types of authorship practices. In the overarching context, it becomes evident that authors whose contributions do not align with established authorship criteria constitute a proportion of 2.56% of the total author population, equating to 14,022 individuals out of a sample of 546,880 authors. While this percentage may appear relatively low, it remains a non-negligible figure within the academic publishing landscape, signifying the importance of addressing these issues.

**Table 3: Number of authors according to types of inappropriate authorship**

| Authorship type | # by type | # by group | % | % by group |
|---|---|---|---|---|
| APC ring | 46 | | 0.01% | |
| Authorship by resources | 5 087 | 14,022 | 0.93% | 2.56% |
| Not meet authorship criteria | 8 889 | | 1.63% | |
| Authorship meets the criteria defined by *PLOS ONE* | 532,858 | 532,858 | 97.44% | 97.44% |
| Total | 546,880 | 546,880 | 100.00% | 100.00% |

When we delve deeper into the specific categories, we find that authors suspected of engaging in the "APC ring" phenomenon represent an extremely minuscule fraction, amounting to just 0.01% of the total author population, involving 46 individuals. This indicates the rarity of such instances within the analyzed publications. In contrast, "authorship by resources" encompasses a significantly larger number of authors, with 5,087 individuals falling into this category.

Lastly, authors failing to meet authorship criteria (outside of financial and resource-related aspects) constitute the largest subset among potential inappropriate authorship practices, accounting for 1.63% of all authors, totaling 8,889 individuals. This underscores the importance of addressing issues related to administrative and supervisory roles that may not warrant full authorship status, necessitating clear delineation between authors and contributors in acknowledgment sections.

**Figure 4: Potential Inappropriate Authorship Concentration Index by country/region**

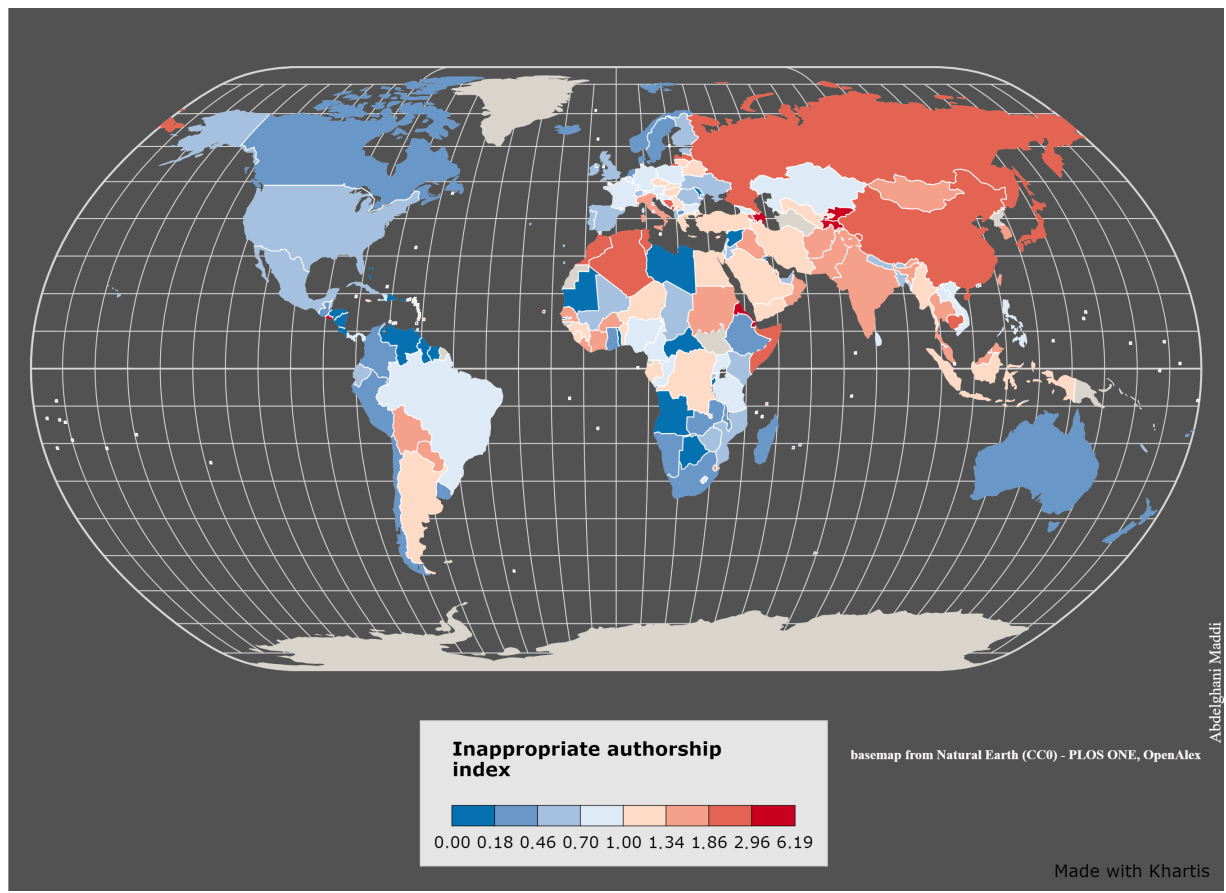

Figure 4 provides insight into the concentration of potential inappropriate authorship practices by country, as measured by PIACI. This index was computed by comparing each country's fractional contribution to the total potential inappropriate authorship with its fractional contribution to the overall *PLOS ONE* publication output. A neutral value for this index is 1. Therefore, if a country has an index of, say, 1.34, it implies that the country's presence in the corpus of publications with potential inappropriate authorship is 34% higher than its presence in the total *PLOS ONE* publication pool.

The findings from Figure 4 are striking, revealing a stark disparity in the concentration of potential inappropriate authorship across different regions. Notably, many countries in Asia and Africa exhibit significantly higher indices, indicating a pronounced overrepresentation in the corpus of publications with potential inappropriate authorship compared to their contributions to the total *PLOS ONE* dataset. This observation suggests that these regions may be more susceptible to issues related to inappropriate authorship practices, warranting further investigation and intervention to uphold ethical authorship standards.

Conversely, Western European countries and North America display notably lower concentration indices, indicating a relatively lower prevalence of inappropriate authorship practices in these regions. These findings underscore the importance of fostering awareness, promoting ethical authorship, and implementing rigorous authorship guidelines, particularly in regions where the concentration of potential inappropriate authorship is more pronounced.

Within the European Union (EU), it is noteworthy that Italy emerges with the highest concentration index in Figure 4, indicating a relatively higher prevalence of potential inappropriate authorship practices within its academic publications. This observation is followed by certain countries in Eastern Europe displaying similar trends. These disparities in potential inappropriate authorship concentrations among EU member states suggest that there may be varying factors at play within different national research ecosystems.

To gain a deeper understanding of these disparities and their underlying causes, further investigation is warranted. Research into the impact of public policies and incentives for publication within these countries could shed light on the proliferation of such phenomena. It is crucial to examine whether certain policies or funding structures inadvertently encourage or fail to deter inappropriate authorship practices. This analysis can help inform evidence-based policy adjustments aimed at fostering ethical research conduct and responsible authorship across the EU and beyond.

Finally, it should be noted that, in Figure 3, we highlighted the strong presence of publications in Ethiopia, which may partly be attributed to multi-collaboration behaviors prompted by the APC discount for LICs. However, in Figure 4, the PIACI does not show a significant increase in Ethiopia. This observation suggests that a combination of analyses, considering both publication volume and authorship attribution dynamics, may yield more nuanced and insightful findings. This underscores the need for further exploration of this phenomenon.

**Table 4: Median total works count and citations number by author, by authorship type**

| Authorship type | Total works count by author (median) | Total number of citations received by author (median) |
|---|---|---|
| Potential APC ring | 104 | 1594 |
| Not meet authorship criteria | 91 | 1316 |
| Potential authorship by resources | 58 | 849 |
| Authorship meets the criteria defined by *PLOS ONE* | 38 | 481 |

Table 4 presents data that could suggest trends in median publication and citation counts across different authorship practices. The data, sourced from the OpenAlex database, indicates a potential pattern in the "Potential APC ring" group. Authors in this category show a median publication count of 104, which might be 2.7 times higher than those adhering to authorship criteria. Similarly, the "APC ring" group exhibits a potentially substantial median citation count of 1,594, compared to 481 in publications with respected authorship criteria.

Efforts were made to analyze authors' academic age, but the available data (as of August 2023) is partial.

In a broader context, the data appears to consistently show higher publication and citation counts in groups not adhering to authorship criteria. This could suggest that authors in these practices might be well-established, with significant publication and citation records. This aligns with expectations, as these authors might typically have better access to resources, contributing to the potential observation of the "Matthew Effect" in bibliometrics. The "Matthew Effect" in bibliometrics refers to the concept that well-known researchers or publications tend to attract more attention, citations, and recognition, leading to a cumulative advantage over time. This suggests that highly cited authors or influential journals are more likely to receive additional citations merely because of their existing prestige. The phenomenon creates a self-reinforcing cycle, where already well-known researchers or works receive even more attention, while lesser-known ones struggle to gain recognition (Merton, 1968; Rossiter, 1993).

In summary, Table 4 hints at the possibility of a prolific nature among authors engaged in certain authorship practices. It suggests a potential influence of established researchers in these practices. The findings underscore the importance of ongoing examination and ethical considerations in scholarly publishing, although conclusions about abuse or misconduct cannot be definitively drawn based on the available data.

**Table 5: Specialization index in inappropriate authorship by affiliation type**

| Affiliation type | Potential APC ring | Potential authorship by resources | Not meet authorship criteria |
|---|---|---|---|
| Other | 2.97 | 0.67 | 0.52 |
| Company | 2.06 | 1.54 | 0.99 |
| Nonprofit | 1.58 | 1.00 | 0.69 |
| Facility | 1.22 | 1.34 | 0.94 |
| Education | 1.01 | 0.87 | 0.99 |
| Healthcare | 0.98 | 1.20 | 1.18 |
| Unknown | 0.39 | 1.17 | 0.96 |
| Government | 0.00 | 1.17 | 1.05 |
| Archive | 0.00 | 1.16 | 0.48 |

**Figure 5: Description of affiliation types presented in Table 5**

```
types*
```
The type of organization based on a controlled list of categories. An organization always has a type. ROR metadata can support multiple types for a given organization, but in most cases, there will just be one type associated with the organization.

Based on the available information about the organization, curators will use their judgement in determining the appropriate category to assign.

Allowed types:

**Education**: A university or similar institution involved in providing education and educating/employing researchers

**Healthcare**: A medical care facility such as hospital or medical clinic. Excludes medical schools, which should be categorized as "Education".

**Company**: A private for-profit corporate entity involved in conducting or sponsoring research.

**Archive**: An organization involved in stewarding research and cultural heritage materials. Includes libraries, museums, and zoos.

**Nonprofit**: A non-profit and non-governmental organization involved in conducting or funding research.

**Government**: An organization that is part of or operated by a national or regional government and that conducts or supports research.

**Facility**: A specialized facility where research takes place, such as a laboratory or telescope or dedicated research area.

**Other**: Use this category for any organization that does not fit the categories above.

Source: https://ror.readme.io/docs/ror-data-structure

Table 5 presents a specialization index (SI) for each type of affiliation, offering additional insights into our previous analysis. This index is derived by calculating a double ratio: for a given type of inappropriate authorship, it relates the proportion of that affiliation type to the proportion of the same type within the total *PLOS ONE* dataset. A neutral value for this indicator is 1. Here, for instance, the "company" affiliation type is 2.06 times more prevalent in the "Potential APC ring" group than in the total *PLOS ONE* dataset.

This table provides compelling supplementary information, highlighting several key findings. It underscores that among authors associated with "Potential APC ring" practices, both companies and non-profit institutions are prominently represented. Similarly, in the case of "Potential authorship by resources," there is a noticeable concentration of "company" affiliations, as well as "facility" affiliations, which encompass institutions providing research facility access such as laboratories and microscopes (as detailed in Figure 5). Lastly, for the third type of inappropriate authorship, "healthcare" affiliations are 18% more prevalent than in the total *PLOS ONE* dataset.

These findings offer a nuanced perspective on our previous results, emphasizing that inappropriate authorship may not necessarily result from misconduct but can be attributed to a misunderstanding of the authorship role. As highlighted by Ali (2021), a primary contributing factor to inappropriate authorship is the lack of knowledge and awareness surrounding authorship guidelines and ethical conduct. Instead of placing financial contributors in the acknowledgments section of articles, some authors may inadvertently include them in the author list. This highlights the importance of clear authorship guidelines and education within the scholarly community to ensure a consistent understanding of authorship standards and practices.

Lastly, it is important to note that the high value of "Other" in the "Potential APC ring" category is attributable to a singular publication in which the corresponding author, who meets the criteria for "Potential APC ring" classification (Marco Vignuzzi), is affiliated with an institution that is inaccurately classified in the OpenAlex database. This publication, accessible via the following link: https://doi.org/10.1371/journal.pone.0252595, constitutes the sole entry in the "Other" category, accounting for 2.33% of the entire APC ring category, while "Other" typically represents 0.79%. Hence, the resulting specialization index attains a value of 2.9.

## 6. Discussion and conclusion

In conclusion, the primary aim of our article was to examine the prevalence of potential inappropriate authorship practices, particularly those stemming from financial or resource-related contributions. To address this objective, we conducted an extensive analysis of publications within the journal *PLOS ONE* spanning the years 2018 to 2023.

Our study has unearthed several insights with profound implications for the landscape of scholarly publishing. Foremost, we have highlighted a notable prevalence of potential inappropriate authorship practices across various categories, signaling the pressing need for institutions and journals to fortify and clarify authorship guidelines. However, several authorities' guidelines are neither uniform nor standard, even within the journals of the same publisher, making the readability and interpretation of such rules difficult (Teixeira da Silva & Dobránszki, 2016). Authorship guidelines should not only outline the criteria for authorship but also provide explicit guidance on acknowledging contributions that may not align with these criteria. This is imperative to safeguard the credibility and integrity of academic publications.

The regional disparities we uncovered in inappropriate authorship practices indicate that certain geographical areas are more susceptible to these issues. Understanding the underlying causes and the influence of local policies on these disparities is essential to promote equitable research participation and ethical authorship standards on a global scale.

Additionally, our analysis revealed that authors engaged in inappropriate authorship practices tend to be prolific and well-established (this concerns the three types of potential inappropriate authorship). This result is different from that of other researchers (Sandler & Russell, 2005; Ni et al., 2021) who showed that non-tenured researchers and women are statistically more likely to be involved in inappropriate authorship. However, the two scenarios can coexist. On the one hand, as highlighted by Khalifa (2022) and Seeman & House (2015), young researchers may be "forced" to include senior researchers (usually their tutors) in their publications as authors. On the other hand, It is also possible, for reasons of competition and the search for stable positions, certain young researchers or women may be involved in inappropriate authorship.

Lastly, the high concentration of companies, nonprofits and facilities among publications exhibiting inappropriate authorship offered a nuanced perspective, emphasizing that inappropriate authorship may not necessarily result from misconduct but can be attributed to a misunderstanding of the authorship role. As highlighted by Ali (2021), the main reason for inappropriate authorship is the lack of awareness among authors of the authorship guidelines. This suggests that in some cases, it is not a question of misconduct *per se*, but rather a lack of vigilance, which can be addressed through training and awareness-raising actions.

These findings carry profound implications for both journal publishers and researchers alike. Journal publishers are prompted to recognize the necessity of implementing stringent control mechanisms to detect and address inappropriate authorship practices effectively. Clear and comprehensive authorship guidelines, coupled with robust editorial oversight, are vital to maintain the credibility and integrity of published research. Researchers, on the other hand, are challenged to discern the appropriate course of action when acknowledging contributors. It becomes imperative for them to differentiate between individuals whose contributions align with authorship criteria and those whose involvement should be duly recognized without inclusion in the author list. By navigating these distinctions conscientiously, researchers contribute to upholding ethical authorship standards and safeguarding the reputation of scholarly publishing.

## 7. Future Research Avenues

In future studies, it would be of particular interest to delve into the phenomenon of inappropriate authorship across different academic disciplines, not only within the confines of *PLOS ONE* but also extending beyond it. This examination could shed light on whether there exist discipline-specific nuances and variations in the prevalence of inappropriate authorship practices. Investigating whether certain fields exhibit a higher propensity for such practices relative to others, including distinctions among the humanities and social sciences, physical and engineering sciences, and life sciences, would provide valuable insights into the interplay of disciplinary norms and ethical authorship standards. Such research could contribute to a more comprehensive understanding of the underlying factors driving inappropriate authorship practices within distinct scholarly domains, facilitating targeted interventions and policy enhancements tailored to specific academic disciplines.

## 8. Limitations

This study has certain limitations that should be taken into consideration. Firstly, it is important to acknowledge that, no matter how rigorous our efforts, it remains impossible to comprehensively verify the true contribution of each author in the publications under scrutiny. Our findings are based on contributions as declared by the authors, which implies a degree of trust in their transparency. Consequently, it is plausible that our figures underestimate the actual extent of inappropriate authorship practices.

On the flip side, it is essential to take into account the potential for overestimation in our data, attributable to *PLOS ONE*'s guidelines stipulating that only corresponding authors are obligated to detail authors' roles during submission. However, we consider this practice advantageous rather than limiting, as relying on corresponding authors enhances the accuracy of declared roles. According to *PLOS ONE*'s author guidelines, the corresponding author is required to indicate "at least one criterion" from CRediT. It is important to note that authors are cognizant of the possibility that non-native English speakers may have misunderstood this directive, which could have impacted the reported statistics (Misra et al., 2018).

Furthermore, it is crucial to note that this study specifically focuses on the case of *PLOS ONE*, one scientific journal among many others. The results obtained in this study cannot be generalized to the entire scientific publishing market. Each journal and research field may have unique authorship practices and characteristics, limiting the scope of our conclusions to this specific case study. Additional research in diverse publishing contexts is necessary to obtain a more comprehensive and nuanced understanding of inappropriate authorship practices in the realm of scientific research.

Smith, E., & Williams-Jones, B. (2012). Authorship and responsibility in health sciences research: a review of procedures for fairly allocating authorship in multi-author studies. *Science and Engineering Ethics*, *18*(2), 199–212. https://doi.org/10.1007/s11948-011-9263-5

Smith, E., Williams-Jones, B., Master, Z., Larivière, V., Sugimoto, C. R., Paul-Hus, A., Shi, M., & Resnik, D. B. (2020). Misconduct and misbehavior related to authorship disagreements in collaborative science. *Science and Engineering Ethics*, *26*(4), 1967–1993. https://doi.org/10.1007/s11948-019-00112-4

Teixeira da Silva, J. A. (2021). Multiple co-first authors, co-corresponding authors and co-supervisors: a synthesis of shared authorship credit. *Online Information Review*, *45*(6), 1116–1130. https://doi.org/10.1108/OIR-06-2020-0219

Teixeira da Silva, J. A. (2023). How are authors' contributions verified in the ICMJE model? *Plant Cell Reports*, *42*(9), 1529–1530. https://doi.org/10.1007/s00299-023-03022-9

Teixeira da Silva, J. A. (2024). The conceptual 'APC ring': Is there a risk of APC-driven guest authorship, and is a change in the culture of the APC needed? *Journal of Scholarly Publishing*, (in press, DOI not yet assigned).

Teixeira da Silva, J. A., & Dobránszki, J. (2016). Multiple authorship in scientific manuscripts: ethical challenges, ghost and guest/gift authorship, and the cultural/disciplinary perspective. *Science and Engineering Ethics*, *22*(5), 1457–1472. https://doi.org/10.1007/s11948-015-9716-3

Teixeira da Silva, J. A., & Rivera, H. (2021). Spousal and kinship co-authorship should be declared to avoid conflicts of interest. *Journal of Bioethical Inquiry*, *18*(3), 379–381. https://doi.org/10.1007/s11673-021-10123-1

Teixeira da Silva, J. A., Tsigaris, P., & Vuong, Q.-H. (2023). Acknowledgments in scientific papers. *Publishing Research Quarterly*. https://doi.org/10.1007/s12109-023-09955-z

Whetstone, D., & Moulaison-Sandy, H. (2020). Quantifying authorship: a comparison of authorship rubrics from five disciplines. *Proceedings of the Association for Information Science and Technology*, *57*(1), e277. https://doi.org/10.1002/pra2.277